\documentclass[12pt,a4paper,twoside]{article}
\usepackage{epsfig}
\usepackage{amssymb,latexsym,amsfonts}

\newcommand{\be}{\begin{equation}}
\newcommand{\ee}{\end{equation}}
\newcommand{\bd}{\begin{displaymath}}
\newcommand{\ed}{\end{displaymath}}
\newcommand{\ba}{\begin{array}}
\newcommand{\ea}{\end{array}}
\newcommand{\bt}{\begin{tabular}}
\newcommand{\et}{\end{tabular}}

\newcommand{\bea}{\begin{eqnarray}}
\newcommand{\eea}{\end{eqnarray}}

\newcommand{\hlf}{\frac{1}{2}}
\newcommand{\qrt}{\frac{1}{4}}
\newcommand{\tqrt}{\frac{3}{4}}

\newcommand{\Z}{\mathbb{Z}}

\newcommand{\R}{\mathbb{R}}

\newcommand{\p}{\mathbb{P}}

\begin{document}
\begin{flushright}
PAR-LPTHE 01-19, LPT-ENS 01/20
\end{flushright}

\vspace{.5cm}

\begin{center}
{\Large Discrete moduli for Type I compactifications}

\vspace{.5cm}

{\bf Arjan Keurentjes}\footnote{email address: \sl arjan@lpthe.jussieu.fr} \\
{\it  LPTHE, Universit\'e Pierre et Marie Curie, Paris VI, Tour 
16,\\ 4 place Jussieu,\\ F-75252 Paris Cedex 05, France  \\

\vspace{.2cm}
 
Laboratoire de Physique Th\'eorique de l'Ecole 
Normale Sup\'erieure, \\24 rue Lhomond,\\ F-75231 Paris Cedex 05, France \\ 
\vskip 0.2cm }
\end{center}

\vspace{.5cm}

\begin{abstract}
We study type I compactification on a 4--torus, with a non-trivial discrete
background RR 4--form field. By using string dualities and recent insights for
gauge theories on tori, we find that a non-trivial background for the RR
4--form is correlated with $Spin(32)/\Z_2$ bundles that are described by a
``non-trivial quadruple'' of holonomies. We also briefly discuss other discrete
moduli for the type I string, and variants of orientifold planes. 
\end{abstract}

\section{Introduction}

The type I string can be regarded as an orientifold of the type IIB string. In
this construction one introduces an $O9^-$ orientifold plane in the theory.
This however causes a tadpole, and consistency requires that this is canceled
by adding 32 D9 branes. The resulting theory has unoriented closed strings,
while the D9-branes introduce an open string sector, leading to a theory with a
gauge group that has as its manifest gauge group $O(32)$. Also a number
of solitons survive the orientifold projection, leading to a spectrum with
D1, D5 and D9 branes. 

Interestingly, some unstable D-brane-anti-brane configurations of type IIB
theory become stable under the orientifold projection: The tachyon arising from
the string states connecting branes and anti-branes is projected out
\cite{Sen98}. Therefore, new non-BPS branes enter the theory. These
non-perturbative objects introduce states in the theory, that transform in
different representations of the gauge group than the states arising from the
ordinary open string sector. It is argued that these fix the topology of the
gauge group to $Spin(32)/\Z_2$, the gauge group of one of the heterotic
theories \cite{Witten98}. Indeed, the existence of these non-perturbative states
is crucial evidence for the conjecture that the type I and the $Spin(32)/\Z_2$
heterotic string are S-dual, and therefore in reality describe two limits of the
same theory \cite{Witten95}.

In type IIB supergravity, the low energy theory to the type IIB superstring,
one encounters a variety of tensorfields, which couple to the extended objects
in the theory \cite{Polchinski}. Introducing the orientifold plane, many of
these fields can no longer fluctuate because that would be incompatible with the
orientifold projection. The fluctuating fields that survive the orientifold
projection are in one-to-one correspondence with the BPS D-branes that appear
in type I theory. The fields whose fluctuations are projected out are
constrained to take constant values over all of space-time. That constant value
does however not necessarily to be equal to zero. It is argued that, before the
orientifold projection, these fields are $U(1)$ valued; the orientifold
projection acts as inversion on the circle which is the group manifold, and
therefore has two fixed points. One can therefore argue that $U(1)$ is broken
to $\Z_2$. The fact that $\Z_2$ is a discrete group demonstrates that
continuous fluctuations are no longer possible, but it leaves a possibility for
non-trivial values for these background fields.

The relevant tensorfields are gauge fields, and the field strengths
corresponding to them must vanish because of the constancy of the potentials.
It is nevertheless possible to construct gauge invariant operators, that can
have non-trivial values if space-time has compact submanifolds. Indeed, the
tensor fields are $n$--forms, and integrating these over compact $n$-cycles
gives us gauge invariant operators. For a 1--form such an operator would
correspond to the standard definition of holonomy. For an $n$--form one has
$(n-1)$ gerbe-holonomy.

A case which has by now been well studied is the possibility for the NS-NS
2--form field $B_2$ to have a non-trivial value over compact 2--cycles
\cite{Bianchi}. This gives a non-trivial phase to closed string world sheets
that wrap around the relevant 2-cycle. Alternatively, one may cut up this
world sheet and interpret it as (a collection of) open string world sheets.
Reproducing the phase factor then places restrictions on the Chan-Paton bundle,
and in fact one can show that the $Spin(32)/\Z_2$ bundle should be
topologically non-trivial \cite{Sen97} (even when the 2--cycle is a 2--torus,
and the bundle is flat \cite{Witten97}). Such bundles have ``absence of vector
structure'' because states transforming in the vector representation of
$Spin(32)$ cannot be consistently introduced in such a background. Of course,
such states are argued to be absent in type I string theory, and the resulting
compactification is consistent.

In the case that the 2--cycle is (topologically) a 2--torus it is possible
to study the bundle \cite{Schweigert}, and consequently the type I string theory
with such a bundle \cite{Witten97} in great detail (see also 
\cite{Kakushadze}). In contrast, there is little known about string 
compactifications with non-trivial backgrounds from the other discrete moduli.

In the present note we will consider the possibility of a non-trivial
background from the RR 4--form. Unlike $B_2$, this does not appear
in perturbative string theory, but it should generate non-perturbative effects.
It would be very interesting to describe compactification on general 4--cycles
allowing one to turn on this background, but we don't know how to do this at
present. For the special case of 4--tori, there are however some recent
developments, that make a study accessible. New insights in constructing flat
bundles on a 3--torus have revealed that the moduli space of flat connections
is much richer than previously thought \cite{Keurentjes99, Kac, Borel}.
Although this research has not (yet) been extended to cover 4-- and higher
dimensional tori, there are some partial results \cite{Kac, Keurentjes00a,
deBoer} that throw sufficient light on the theory we are interested in, the
$Spin(32)/\Z_2$ gauge theory appearing in the type I string. Armed with these,
and string dualities we will describe type I compactifications on a 4--torus,
with non-trivial background from the RR 4--form. On the 4-torus one can of
course also turn on a non-trivial NS-NS 2--form, and we will describe also these
compactifications.

\section{D-branes in background NS-NS and RR-fields}

We will briefly review some aspects of toroidal compactification with
non-trivial holonomy for the NS-NS 2--form, and point out some parallels and
differences with the case we are interested in, holonomy for the RR 4--form.

For the NS 2--form, it is convenient to first study its effect in generality,
and only afterwards introduce orientifold planes. A very simple way to study
what happens when one turns on a non-zero $B_2$-field over some 2-cycles of an
$n$-torus is by using T-duality. Under T-duality the metric and $B_2$-field
moduli mix (see \cite{Giveon} for a review), and the dual torus has angles
different from the original one. Consider a square 2-torus, with an
appropriately normalized $B_2$-field. Let there also be a D-brane wrapping the
2-torus (possibly multiple times) and suppose there is a field strength $F_2$
present on the brane. Set 
\bd \int_{T^2} B_2 =b; \qquad \int_{T^2} F_2 =f . \ed
Let there be a single D-brane wrapping the 2-torus (possibly multiple times)
Applying a single T-duality, the dual torus becomes a skew one. The angle
$\phi$ between two basis vectors for the lattice for this torus is given by
$\tan \phi = b$. The single D-brane is dualized to a brane wrapping one cycle
of the torus. This brane makes an angle $\psi$, given by $\tan \psi = -f$ with
the other cycle. Closure of the D-brane now translates into the condition 
\be
n(b+f) \in \Z \quad \textrm{for } n \in \Z.
\ee
The number $n$ is appropriately interpreted as ``wrapping number'' of the
D-brane. This obviously implies that $b+f$ is a rational number. Solutions with
$f=0$ exist if and only if $b$ is a rational number. Assume this to be the case
and let $m$ be the smallest integer such that $mb \in \Z$. Then a D-brane
wrapping the dual torus $m$-times, with $\psi =0$ describes the dual theory to
a D-brane wrapping the original torus multiple times, which has on its
world volume a gauge theory which is described by an $U(m)$ bundle with twisted
boundary conditions. These boundary conditions break the $U(m)$ to $U(1)$,
justifying in hindsight our ignoring of the non-Abelian interactions. As a side
remark, we note that in the general case, where we wrap the brane multiple
times and introduce a non-zero field strength, one can proceed by decomposing
$U(m)$ as $(U(1) \times SU(m))/\Z_m$. One then uses twisted boundary conditions
in $SU(m)$, and puts the field strength in the $U(1)$ factor. Again, this breaks
the effective gauge group to $U(1)$, and our ignoring of the non-Abelian
interactions is justified. 

By T-duality the conclusions from the previous paragraph can be translated
back to the original theory. When one wraps a brane over a  2--torus with
non-trivial $B_2$-holonomy one should compensate for the effects by turning on
a field strength on the brane, and/or wrapping the brane multiple times. The
same conclusion can be reached from a more advanced argument. Consider a
string world sheet ending on the D-brane, wrapped around a 2--cycle. In the
path-integral there appears a phase-factor \cite{FW}
\be 
\exp \left(i \int_\Sigma B + i \oint_{\partial \Sigma} A \right).
\ee 
In \cite{FW} an additional factor was considered, coming from the Pfaffian of
the Dirac operator on the world sheet, but this is not relevant to our present
considerations. For reasons explained below, the total factor should be equal to
unity. Of course $\oint_{\partial \Sigma} A$ can simply be converted into
$\int_{ \Sigma} F_2$. Then, trivializing the phase factor requires turning on an
appropriate field strength, and/or wrapping the brane multiple times over the
cycle. This extends the previous result to generic 2--cycles.

In a somewhat more specific context, the same argument was already used
in \cite{Sen97}. Here type I theory on a K3 was considered, where the
2-cycles have the topology of spheres. Type I theory only allows $B_2$-fields that are
multiples of $\hlf$. Turning on a non-integer $B_2$-field is correlated with a
choice of Chan-Paton bundle over the 2-cycle. To be precise, the bundle should
be one without vector structure. The interpretation of the authors of
\cite{Sen97} is that the correlation between $B_2$-field and Chan-Paton bundle
is required by consistent coupling of closed strings to open strings. The
anomaly of \cite{FW} can be interpreted in a similar way: trivializing the
phase factor is necessary to couple the open strings ending on the D-brane to
closed strings living in the bulk. If the factor cannot be trivialized, then
the only option left is to remove the particular open string sector; in other
words, to discard the possibility to wrap the brane around the 2-cycle. This
truncation of the spectrum however may lead to other inconsistencies. For
example, in the presence of orientifold planes D-branes are needed to ensure
tadpole cancellation.

In this fashion the rank reduction in the case of toroidal compactification
without vector structure can be understood as follows. In type I theory on a
torus a half-integer $B_2$-field gives a phase factor that can be
canceled in two ways: One can turn on a non-zero gauge field strength,
or wrap the D9 branes twice around the corresponding cycle. It is clear
that wrapping twice gives a solution that is lower in energy than the
turning on of a field strength. However, locally a twice wrapped D-brane is
indistinguishable from two once wrapped D-branes. In particular, we need only
16 D9-branes to cancel the $O9$ tadpole, instead of the usual 32.

Absence of vector structure for a $Spin(32)/\Z_2$ bundle is measured by a
characteristic class, a $\Z_2$-valued generalized Stieffel-Whitney class
$\tilde{w}_2$. As the choice of bundle is correlated with the choice of
$B_2$-field, it follows that one may identify the $B_2$-field with $\tilde{w}_2$.
It is an interesting observation that discrete moduli for string theory are
correlated with topological invariants of particular gauge bundles. There also
appear to be suggestive links between the 3--form in M-theory, and the
Chern-Simons 3--form associated to particular bundles \cite{Diaconescu}
\cite{deBoer}.

In this paper we wish to describe type I on a 4--torus with a RR 4--form field
turned on. As we will see, a non-trivial RR 4--form field indeed results in
reduction of the rank of the gauge group. This is however not due to multiply
winding branes. Na\"\i vely applying dualities to the reasoning of \cite{FW}
seems to imply that also a RR 4--form flux has consequences for brane
wrappings. This is not the case for the following reason: The NS-NS $B_2$-field
couples to strings, and the reasoning of \cite{FW} is in terms of string
world sheets. Dualizing the set-up, this says that the pullback of the RR
4--form field to any 3-brane world volume intersecting the D-brane must vanish.
Type I theory however has no 3-branes, not even non-BPS ones \cite{Witten98}.
Therefore we should expect the same number of once wrapped branes,  regardless
of whether the RR 4--form field is turned on or not. Why one nevertheless gets
reduction of the rank will be explained in the next section.

Another interesting question is whether the discrete 4--form of type I theory
can be identified with some topological invariant of the $Spin(32)/\Z_2$
bundle. A piece of evidence is the existence of flat non-trivial bundles that
are intrinsically 4-dimensional \cite{Kac}. In appendix D of this paper
``non-trivial quadruples'' were constructed, 4--tuples of group elements such
that every subset of these 4 elements can be chosen on a maximal torus
of the group, but not all four simultaneously. Such a 4--tuple can be used for a
compactification on the 4-torus, by choosing the elements of the 4--tuple as
holonomies. 

Bundles over the 4--torus parametrized by such a 4--tuple are intrinsically
4-dimensional (as the bundle over every sub 3-torus is a standard
compactification). One may speculate on the existence of a 4-dimensional
topological invariant, distinguishing the compactification with a non-trivial
quadruple form a trivial one. Unfortunately, such an invariant has not (yet)
been constructed. This is to be contrasted with compactifications of gauge
theories on lower dimensional tori, where the invariants classifying the
bundles are understood, and do allow generalization to non-toroidal
compactifications. We will argue that type I string theory has a candidate for
such an invariant, provided by the RR 4--form. We will show that turning on this
form reproduces the non-trivial quadruple of \cite{Kac}. It is also possible to
combine non-zero expectation values for both the $B_2$-field and the RR
4--form, leading to a ``non-trivial quadruple without vector structure'' as
first encountered in \cite{deBoer}.

We note that the fact that we are not going to find multiply wrapped branes can
also be heuristically understood from the link to bundles in gauge theories.
Non-trivial quadruples in the sense of \cite{Kac} only exist in (large enough)
orthogonal groups. When extending to ``c-quadruples'' (which we define
extrapolating on the definitions of \cite{Borel}, as 4--tuples of elements
which commute up to elements of the center of the simply connected cover of the
group), it is possible to show that these may also appear in theories with
symplectic groups. Important however is, that it is simple to show that they do
not occur for the unitary groups. Translated to string theory this suggests
that the orientifold projection is essential to the effects of the RR 4--form,
at least in the context that we wish to study. Of course this makes the effect
of the RR 4--form in absence of an orientifold projection even less understood.
 
\section{Identification via T-duality}

Consider type I on a 4-torus, with possibly some $B_2$-fields, and possibly the
RR 4--form turned on. There are, up to $SL(4, \Z)$ transformations, three
possibilities for $B_2$-flux over the 4--torus, that can be distinguished as
follows. Viewing $B_2$ as a 2--form with integer periods, one has the
possibilities \bea 
B_2 =0; & B_2 \neq 0, B_2^2 = 0; &  B_2 \neq 0, B_2^2 \neq 0. \label{bf}
\eea
These are of course in precise correspondence with the topological choices for
the $Spin(32)/\Z_2$ bundle over the 4--torus \cite{Witten97}. 

On the 4-torus, a 4--form is invariant under $SL(4,\Z)$ transformations (because
it has to be proportional to the volume form). Calling the 4--form $C_4$, there
are 2 possibilities
\bea
C_4=0; & & C_4 \neq 0.
\eea
Hence a priori there are 6 possibilities to consider.

Applying 4 T-dualities to the type I theory on the 4--torus gives us a IIB
orientifold on $T^4/\Z_2$ with 16 $O5$ planes at the fixed points of the $\Z_2$
action. The identities of the $O5$-planes depend on the fluxes in the parent
type I model. The 4 T-dualities result in a $B_2$-field background that is
identical to that of the parent model on $T^4$.

There are 2 possible discrete charges for $O5$ planes: its transverse space is
$\R \p^3$, which has the following interesting cohomologies. There is a
possibility for a discrete charge for the NS-NS 2--form, as its class
\bea
[dB_2] = [H_3] & \in & \tilde{H}^3(\R \p^3) = \Z_2.
\eea
Another possibility is a discrete charge for the RR-scalar, as 
\bea
[dC_0] = [G_1] & \in & \tilde{H}^1 (\R \p^3) = \Z_2.
\eea
By these two $\Z_2$ charges, it is possible to distinguish 4 types of O5-planes
which we will denote as $O5^-$, $O5^+$, $\widetilde{O5^-}$ and
$\widetilde{O5^+}$. Here we follow the notation of \cite{Hanany}, using the
superscript $+$ for planes with non-trivial NS-charge, and a tilde for planes
with non-trivial RR charge. The D5 brane charges of these orientifold planes
are $-2$ for the $O5^-$, $-1$ for the $\widetilde{O5^-}$,  and $+2$ for
the $O5^+$, $\widetilde{O5^+}$. The difference in charge between $O5^-$ and
$\widetilde{O5^-}$ gives rise to the interpretation of the second as a bound
state of an $O5^-$ with a single D5-brane. The D5-brane charge of $O5^+$
and $\widetilde{O5^+}$ is the same, and the two can presumably only be
distinguished by non-perturbative effects. Nevertheless, we will find that both
make a natural appearance.

The distribution of the NS-charges over the $O5$ planes in the IIB orientifold
follows from the $B_2$-fluxes in the original type I model\cite{Witten97}.
The 3 cases give  \begin{table}[h] \begin{center}
\begin{tabular}{|c||c|c|c|}
\hline
$B_2$-flux in type I on $T^4$ &
$B_2 =0$ & $B_2 \neq 0$, $B_2^2 = 0$ &  $B_2 \neq 0$, $B_2^2 \neq 0$\\
\hline
planes with non-trivial B-charge & 0 & 4 & 6 \\
\hline
\end{tabular} \end{center}
\end{table}

In the case of 4 planes with non-trivial NS-charge, the intersection points
of these 4 $O5$-planes are aligned within a 2-plane. In the case of 6 planes
with non-trivial charges, the 6 planes are aligned in two 2-planes intersecting
in a point, with a plane with trivial NS-charge at the intersection point.

The RR 4--form of the type I theory on the 4--torus results after 4 T-dualities
in a constant RR scalar background for the IIB orientifold on $T^4/\Z_2$. This
inevitably implies that all orientifold planes have the same RR-scalar charge,
as it can be measured at any point near the $O5$-plane. 

In a recent paper \cite{Hyakutake} also the possibility of gradients for the
RR-scalar between $O5$-planes was considered. Such a gradient gives a non-zero
field strength, and via coupling to gravity should modify the curvature of
space-time. Solutions to the combined problem (solving the Einstein equations,
for a space with some compact directions, and with appropriate symmetries such
that the orientifold planes can be inserted) would be very interesting, but
probably break some supersymmetry, and it seems unlikely that they are related
to the type I string in a simple way. We will therefore discard this
possibility.

The 6 possibilities for the duals to the type I string compactification
on a 4-torus have the following configurations of $O5$-planes:
\begin{itemize} 
\item All 16 planes are $O5^-$ planes. This is the trivial compactification, and
there are 32 D5-branes, arranged in 16 pairs, giving a rank 16 gauge group.
\item 4 $O5^+$ planes and 12 $O5^-$ planes. This is dual to the compactification
without vector structure discussed in \cite{Witten97}. There are 8 pairs
of D5-branes.
\item 6 $O5^+$ planes and 10 $O5^-$ planes. This is another compactification
that was briefly described in \cite{Witten97}. There are 4 pairs of D5-branes.
\item All 16 planes are $\widetilde{O5^-}$ planes. This compactification is
dual to a type I compactification with a non-trivial quadruple. It is trivial
to reconstruct the holonomies for this case, and compare them with \cite{Kac}.
Note that there are 32 D5-branes, of which 16 form bound states with the
$O5$ planes, while 16 others are arranged in pairs. Therefore the rank of the
gauge group is 8, and, remarkably, it can actually be demonstrated that this
orientifold is in the same moduli space as the one with 4 $O5^+$ planes and 12
$O5^-$ planes \cite{deBoer}. 
\item 4 $\widetilde{O5^+}$ planes and 12 $\widetilde{O5^-}$ planes. This model
only appeared previously briefly in \cite{deBoer}. It has 16 D-branes
like the standard compactification without vector structure, of which 12 are
bound to $O5$ planes, and 4 are arranged in pairs. Therefore the rank of the
gauge group is only 2. In \cite{deBoer} it was conjectured to be dual to the
type I compactification with a quadruple without vector structure. In the
next section we will give compelling evidence for this duality by
reconstructing the Wilson lines from this orientifold.
\item The last model would have 6 $\widetilde{O5^+}$ planes and 10
$\widetilde{O5^-}$ planes. Now the tadpole can only be canceled by
adding anti D5-branes in pairs, which should then annihilate with the single
D-branes stuck to the $O5$ plane, resulting in a model where some of the $O5$
planes form bound states with an anti D5-brane. Actually, if this is a valid
possibility, then it is impossible to tell which $O5$ planes have the anti
D5-branes, because all possibilities can be realized. In fact, they
should be realized, because by D5 brane-anti brane pair creation in the bulk,
and letting these annihilate with branes and anti-branes bound to $O5$-planes,
the anti D5-brane can ``jump'' to other $O5$-planes. The true ground state of
the theory would then be a superposition of all possible configurations. We
will not study this non-supersymmetric model further. Incidentally, we note
that a model with 6 $\widetilde{O5^+}$ planes and 10 $\widetilde{O5^-}$ planes
describes a perfectly sensible gauge bundle for $Spin(N \geq  40)$ gauge
theory (compare with \cite{Keurentjes00b}). It is just the fact that the
(perturbative) gauge group of type I is ``not big enough'' that leads to the
subtleties mentioned. \end{itemize}

For completeness we mention that there exist two more IIB orientifolds on
$T^4/\Z_2$ (with 16 supersymmetries), both with 8 $O5^+$ planes and 8 $O5^-$
planes. It was noted in \cite{deBoer, Bergman} that there already exist
two geometrically inequivalent configurations for IIA on $T^3/\Z_2$ with 4
$O6^+$ and 4 $O6^-$ planes. Compactifying these on a circle and T-dualizing
leads to two inequivalent configurations on $T^4/\Z_2$. It is clear however
that for these cases we cannot add an RR-scalar background, as this would lead
to tadpoles or supersymmetry breaking. 

The existence of inequivalent theories with an equal number of $Op^+$ and $Op^-$
planes allows an elegant explanation. In \cite{Witten97} it is argued that these
theories have their origin in a special orientifold of IIB theory on a circle.
Instead of orientifolding this theory straight away, the orientifold action
$\Omega$ is combined with half a translation $\delta$ over the circle. Another,
equivalent way of stating this is that it is IIB on a circle with a special
holonomy around the circle \cite{Keurentjes00b}. Just like in the case
compactifications of the ``ordinary'' orientifold, which leads to type I
theory, compactifications of this ``special'' orientifold of type IIB theory
allow a non-trivial background $B_2$ field, but it can only take two discrete
values. Compactifying IIB on a circle, modded by $\delta \Omega$ on a
further 2--torus, one has the choice of turning on a discrete $B_2$ field.
Therefore in 7-d, there are 2 inequivalent theories, which after dualizing
result in the 2 inequivalent IIA orientifolds with 4 $O6^+$ planes and 4 $O6^-$
planes. In the same fashion, one immediately sees that the IIA orientifold on
$T^5/\Z_2$ with 16 $O4^+$ and 16 $O4^-$ allows 3 inequivalent geometries (from
the 3 inequivalent choices of $B_2$-field, as in eq. (\ref{bf})), the IIA
orientifold on $T^7/\Z_2$ with 64 $O2^+$ and 64 $O2^-$ comes with 4
inequivalent geometries, and IIA on $T^9/\Z_2$  with 256 $O0^+$ and 256 $O0^-$
exists in 5 inequivalent geometries (more about this in \cite{Keurentjes01}). We
also note that these considerations make a non-perturbative equivalence of the
different configurations with equal numbers of $Op^+$ and $Op^-$, mentioned in
\cite{deBoer} but conjectured to be false, indeed highly unlikely. 

In total, we have identified 7 orientifolds on $T^4/\Z_2$ preserving
16 supersymmetries (It can actually be shown that these are all maximally
supersymmetric orientifolds with 16 O5-planes only \cite{Keurentjes01}).

\section{The model with $\widetilde{O5^+}$ planes}

In this section we will take a closer look at the model on $T^4/\Z_2$ with 4
$\widetilde{O5^+}$ planes and 12 $\widetilde{O5^-}$ planes. This appears to be
the only supersymmetric model in which $\widetilde{O5^+}$ planes quite
naturally appear. $\widetilde{Op^+}$ planes with $p < 5$ have been studied
before \cite{Hanany, Hori, Gimon}. 

Our explanation for the appearance of the $\widetilde{O5^+}$ originates in the
$B_2$ and $C_4$ holonomies appearing in the type I model. The NS-charge of the
$\widetilde{O5^+}$  can also be confirmed by studying the low energy gauge
theory, as it is supposed to lead to $Sp(n)$ gauge symmetry. The RR-charge can
not be confirmed this way, as both $\widetilde{O5^+}$ and $O5^+$ lead to the
same low energy gauge group, $Sp(n)$. But there is another piece of evidence
that (we think) supports our assignment of charges.

In \cite{deBoer} the 2 models with 4 $O5^+$ planes and 12 $O5^-$ planes, and 16
$\widetilde{O5^-}$ planes, where shown to be in the same moduli space. The
most unambiguous way to show this is to translate both models to heterotic
string theories, whose equivalence can be demonstrated exactly \cite{deBoer}.
Another, less precise way to exhibit the close relationship between the two
models is to compactify both on an additional 2--torus, and T-dualize along the
2 directions of this torus. This leads to IIB orientifolds on $T^6/\Z_2$ with
16 $O3^+$ planes and 48 $O5^-$ planes, and another with 16 $\widetilde{O3^-}$
planes and 48 $O3^-$ planes. These are dual to each other by S-duality of 4-d
$N=4$ supersymmetric gauge theories, which is realized as a $\Z_2$ involution on
the component of the string moduli space that contains the CHL-string, as well
as the above two models \cite{Chaudhuri, Mikhailov, deBoer}.

It is interesting to apply the same procedure to the model with 4
$\widetilde{O5^+}$ planes and 12 $\widetilde{O5^-}$ planes. Compactification on
a 2--torus, and applying T-dualities twice results in a model on $T^6/\Z_2$
with 4 $\widetilde{O5^+}$ planes, 12 $\widetilde{O5^-}$ planes, 12 $O5^+$
planes and 36 $O5^-$ planes. This model is self-dual under S-duality of 4-d
$N=4$ supersymmetric gauge theories ! Indeed, in \cite{deBoer} this model was
conjectured to be dual to the $\Z_4$ triple construction in the $E_8 \times
E_8$ heterotic string. The component of the string moduli space that contains
these theories is mapped to itself under the $\Z_2$ involution implied by
S-duality of 4-d $N=4$ theories. The self-duality under 4-d S-duality of the
compactification to 4-d of the orientifold with 4 $\widetilde{O5^+}$ planes and
12 $\widetilde{O5^-}$ planes is clearly consistent with the other proposed
dualities.

Although in principle the RR-charges of the $Op^+$ planes could also be
determined by studying the monopole spectrum in $d=3$ \cite{Hanany}, this is
very subtle. The reason for this is that there are only 2 D-brane pairs present
in the theory, and hence the only gauge groups one can get at $Op^+$ planes are
$Sp(1)$ and $Sp(2)$. Therefore the groups of the monopole theory can only be
$SO(5) = Sp(2)/\Z_2$ and $SO(3) = Sp(1)/\Z_2$. Hence, a determination of the
gauge groups appearing in the S-dual theories is not enough, one really needs to
study the topology of the gauge groups in detail. Together with the fact that
in type I theory and its duals the topology of the gauge group is different
from the one that is manifest in perturbation theory \cite{Witten98} (Note also
the issues raised on the topology of the gauge group in \cite{deBoer}), the
analysis appears to be very difficult, and not necessarily decisive. We will
not attempt such an analysis here.

We will now reconstruct the same model in another way, which clarifies
the duality to the $\Z_4$ triple in $E_8 \times E_8$ heterotic string theory.

We start again with type I string theory on a 4--torus. Start by turning on a
Wilson line on the first circle which breaks the gauge group to a group that
can be shown to be $(Spin(16) \times Spin(16))/\Z_2$ (see \cite{deBoer} for a
discussion on the topology of subgroups in string theory). This group is the one
that is often denoted as ``$SO(16) \times SO(16)$'', for
example in discussions on T-duality of heterotic theories. T-dualizing
along this direction now leads to a model where one half of the D-branes is
localized at $x_1 = 0$ and the other half is at $x_1 = \pi$ (in an obvious
choice of coordinates). 

The point is that $Spin(16)/\Z_2$ allows various triples without
vector structure \cite{Borel}. There are 4 distinct possibilities. An example
of the first kind combines 2 holonomies parametrizing absence of vector
structure, while taking for the third holonomy the identity. By continuous
deformations one can reach other models on the same component of the moduli
space. The Chern-Simons invariant for these triples is integer \cite{Borel}.
This is the only triple without vector structure that is allowed in string
theory \cite{deBoer}. The other ones are nevertheless useful as building
blocks, as we will demonstrate.

A second example combines 2 holonomies parametrizing absence of vector
structure, with a third holonomy corresponding to the non-trivial element of
the center of $Spin(16)/\Z_2$. Again the rest of this component in the moduli
space can be covered by continuous deformations. This component has
Chern-Simons invariant equal to $\hlf$ plus some integer, and we will not
discuss it further.

The third and fourth example are more interesting. By choosing an appropriate
Wilson line, it is possible to break $Spin(16)/\Z_2$ to $(SU(4) \times
Spin(10))/\Z_4$ (with the $\Z_4$ acting diagonally on both factors). Choosing
this element as our third holonomy, and two other holonomies that commute up to
an element of order 4 in the $\Z_4$ that was divided out of $SU(4) \times
Spin(10)$, one finds 3 holonomies that commute in $Spin(16)/\Z_2$. They do not
commute when lifted to $Spin(16)$ and therefore define a triple without vector
structure. There are basically two options (because there are 2 elements of
order 4 in $\Z_4$) leading to Chern-Simons invariants of $\qrt$ and $\tqrt$ up
to integers. These models do allow an orientifold description, which was
constructed in \cite{Keurentjes00b} (section 4.3.2).

Compactifications with non-integer Chern-Simons invariants are not allowed in
consistent string theories \cite{deBoer}. Here however, we have the group
$(Spin(16) \times Spin(16))/\Z_2$, which allows us to embed a triple without
vector structure in each $Spin(16)$ factor. Choosing the triple with
Chern-Simons invariant $\qrt$ in one factor, and the triple with Chern-Simons
$\tqrt$ in the other, all possible Chern-Simons invariants that can be defined
over the 4--torus are integer, and these holonomies define a consistent string
background. This also illustrates the equivalence of the present construction to
the formulation of the $E_8 \times E_8$ $\Z_4$-triple theory, that appears in
\cite{deBoer}. Finally, using the (inconsistent) orientifolds parametrizing the
triple theories that were constructed in \cite{Keurentjes00b}, one can
trivially construct the (consistent) orientifold representation of the present
quadruple without vector structure: We had half of our D-branes living at $x_1
= 0$ and one half at $x_1 = \pi$, and copying the orientifold description for
the triple theories leads immediately to an orientifold on $T^4/\Z_2$ with 12
$\widetilde{O5^-}$ planes. Furthermore there are 4 planes that are either
$\widetilde{O5^+}$ or $O5^+$ planes. This cannot be decided from an analysis of
the gauge group, but we have given other evidence that these actually should
be $\widetilde{O5^+}$ for a consistent string theory.

To complete this section, we will deduce the holonomies parametrized by the
IIB orientifold on $T^4/\Z_2$ with 4 $\widetilde{O5^+}$ and 12
$\widetilde{O5^-}$ planes. One can use the techniques described in
\cite{Keurentjes00b}. We start by first defining some building blocks 
\be  A = \left( \ba{rr} 1 & 0 \\ 0 & -1 \ea \right) \qquad B = \left( \ba{rr}
0 & -1 \\ 1 & 0 \ea \right) \qquad C =\left( \ba{rr} 0 & 1 \\ 1 & 0 \ea \right)
\ee \be D(\phi) = \left( \ba{rr} \cos \phi & -\sin \phi \\ \sin \phi & \cos
\phi \ea \right) \ee 
The holonomies can then be expressed in a relatively
compact way as \bea
\Omega_1 & = & (A \oplus B \oplus C)^4 \oplus (D(\phi_1) \otimes A) \oplus
(D(\psi_1) \otimes A) \\
\Omega_2 & = & (B \oplus A \oplus A)^4 \oplus (D(\phi_2) \otimes C) \oplus
(D(\psi_2) \otimes C) \\
\Omega_3 & = & \textrm{diag} (1^{12}, (-1)^{12}) \oplus (D(\phi_3)^2) \oplus
(D(\psi_3)^2) \\
\Omega_4 & = & \textrm{diag} (1^{6}, (-1)^{6}, 1^{6}, (-1)^{6}) \oplus
(D(\phi_4)^2) \oplus (D(\psi_4)^2) \eea
The notation with superscripts indicates that the corresponding arguments have
to be repeated. The arguments $\phi_i$ and $\psi_i$ are the coordinates of
the 2 pairs of D-branes on the orientifold, suitably normalized. The reader may
verify that $\Omega_1$ and $\Omega_2$ anticommute, that all other combinations
of holonomies commute, and that the eigenvalues of the holonomies coincide with
the ones given for the heterotic $Spin(32)/\Z_2$ with a quadruple without
vector structure as given in \cite{deBoer}, translated to the vector
representation of $Spin(32)$.

\section{Other moduli ?}

At least na\"\i vely, the reasoning that leads to considering the
possibility for non-trivial discrete NS 2--form and RR 4--form backgrounds in
the type I string, seems to suggest still more possibilities for discrete
moduli. In principle, one may also study type I theory on a sufficiently large
torus, with non-trivial backgrounds for the RR 8--form and the NS
6--form. After applying T-dualities, this presumably leads to the additional
variants for orientifold lines and points considered in \cite{Hanany}
(compactifications with non-trivial NS 6--form background are currently under
study \cite{Sav}).

In view of the above it is natural to ask whether one can turn on a background
for the RR-scalar. After T-dualities this would lead to RR $p$--form charges for
$O(9-p)$ planes. These charges do not appear in \cite{Hanany}. This is because
such fluxes would fill the whole transverse space to the orientifold plane, and
in particular can not be studied with the cohomologies of the $\R \p^{8-p}$ that
surrounds the $Op$-plane. This is not necessarily an obstruction to their
existence, as the same problem exists for $O8^+$ and $O7^+$ planes, and can be
overcome (see e.g. \cite{Witten97} \cite{deBoer}). Incidentally we note that
this may imply that $Op$ planes with $B_6$ fluxes 
may exist in dimensions higher than 1.

At least in some cases it seems to be possible to make sense out of orientifold
$Op$ planes with $C_{9-p}$-flux (where $C_{9-p}$ denotes a transverse RR
$9-p$--form flux). The first example that comes to mind is by using type IIB
S-duality on a $O7^+$ plane. In this case one can easily determine the charges,
the tension and the gauge group associated to the resulting plane.

As a second example we consider $C_1$-flux for $O8$-planes. There exist
variants of $O4$ planes and $O0$-points that carry a similar flux. When lifting
these planes to M-theory, $C_1$ becomes a component of the metric. In
particular, it can be demonstrated that the $O4$ and $O0$ lift to M-theory on
\cite{Hori, Gimon, Hanany} 
\bd
\R^{4,1} \times (\R^5 \times S^1)/\Z_2 \qquad \textrm {and} \qquad \R^{0,1}
\times (\R^9 \times S^1)/\Z_2
\ed
Here the $\Z_2$ acts as a shift on $S^1$ and as a reflection on $R^n$. In full
analogy, an $O8$ with non-trivial $C_1$ charge should lift to M-theory on
\bd
\R^{8,1} \times (\R^1 \times S^1)/\Z_2 
\ed
But in that case, this object is already known, as this description applies
(locally) to M-theory on a Klein bottle \cite{Dabholkar}, and on a M\"obius
strip. Such an $O8$-plane would carry D8 brane charge 0.

A third (more speculative) example is provided by $C_3$-flux for $O6$ planes.
In \cite{deBoer}, M-theory on a K3 with background 3--form fluxes was
considered. For a background $\Z_2$ valued 3--form flux, one needs an even
number of frozen $D_4$ singularities. If one could split of a circular fiber,
in such a way that pairs of $D_4$ singularities are located in the same fiber,
then presumably the theory can be reduced to a IIA theory, with a 3--form flux
background. Comparing various charges, the fiber with two $D_4$ singularities
reduces to an object with D6 brane charge 12, and a non-trivial 3--form charge.
Lifting a single object in IIA theory to two singularities may seem unusual,
but the reader may wish to compare with the case of the $O6^-$ with 4 D6-branes
on top, that lifts to two $A_1$ singularities.

There seems to be a pattern consisting of $Op$-planes, with Dp brane charge
$16-2^{p-4}$, and RR $(9-p)$--form charge. Is it possible to have an $O9$ plane
with these charges ? For several reasons, this is problematic. First of all,
such an $O9$-plane could be used to construct a new 10 dimensional open string
theory with a rank 8 gauge group, but this theory is not known. Second, this
theory would appear (via M-theory on the M\"obius strip) to be a 10 dimensional
limit of the CHL-string, but various arguments (see e.g. \cite{Mikhailov})
indicate that such a limit does not exist. This suggests that the discrete
$C_0$ flux can only be defined on a manifold with compact directions. It would
be interesting to investigate this possibility further.

A serious drawback of the previous considerations, is that we more or less
``define'' variants of orientifold planes by projections from non-perturbative
descriptions (type IIB $SL(2, \Z)$ duality, M-theory). This is opposite to
common practice, where the perturbative objects are well defined,
and one tries to deduce the description in the strong-coupling regime. To some
extent, the question is whether these variants of orientifold planes (and the
ones introduced in \cite{Hanany}) are really sensible as string theory objects,
or whether they only start to make sense in a more complete, non-perturbative
description of string theory.

\section{Conclusions}

We have demonstrated that type I compactifications on a 4--torus with a
non-trivial RR 4--form background field lead to theories with gauge groups of
reduced rank. The RR 4--form was shown to be correlated with compactifications
with ``non-trivial quadruples'' of holonomies. It is not known at present
whether there exists a characterization of these bundles that extends to other 4
manifolds. For example, it would be very interesting to compactify type I
theories on K3, or a Calabi-Yau manifold with 4--cycles, and turn on the
discrete 4--form background, but at present we have no clue how to describe
such compactifications. 

In one of the present models, the $\widetilde{O5^+}$ makes a natural appearance.
The existence of this plane was deduced before, from an analysis of possible
discrete charges, but it has not appeared thus far in an explicit model.

We briefly discussed some aspects of variants of orientifold planes. The fact
that it appears to be much easier to define these objects from M-theory or
using S-dualities in IIB theory, warrants the question whether they are really
well defined in perturbative string theory. Instead it does not seem unlikely
that only a non-perturbative description reveals all the properties of these
planes. 

{\bf Acknowledgements}:  We would like to thank J. de
Boer, R. Dijkgraaf, A. Hanany, D. Morrison, and S. Sethi for helpful
conversations. This work is partly supported by EU-contact HPRN-CT-2000-00122.

\end{document}